\documentclass[12pt,preprint]{aastex}
\usepackage{underscore}
\usepackage{color}

\slugcomment{Accepted on 2014 December 11 for publication in the Astrophysical Journal Letters}

\shorttitle{209P/LINEAR} 
\shortauthors{Ishiguro et al.}

\begin{document}

\title{Dust from Comet 209P/LINEAR during its 2014 Return:\\
Parent Body of a New Meteor Shower, the May Camelopardalids
}

\author{Masateru \textsc{Ishiguro}\altaffilmark{\dag}}
\affil{Department of Physics and Astronomy, Seoul National University,
Gwanak, Seoul 151-742, South Korea}

\author{Daisuke \textsc{Kuroda}}
\affil{Okayama Astrophysical Observatory, National Astronomical
Observatory of Japan, Asakuchi, Okayama 719-0232, Japan}

\author{Hidekazu \textsc{Hanayama}}
\affil{Ishigakijima Astronomical Observatory, National Astronomical
Observatory of Japan, Ishigaki, Okinawa 907-0024, Japan}

\author{Jun \textsc{Takahashi}}
\affil{Nishi-Harima Astronomical Observatory, Center for Astronomy,
University of Hyogo, Sayo, Hyogo 679-5313, Japan}

\author{Sunao \textsc{Hasegawa}, Yuki \textsc{Sarugaku}}
\affil{Institute of Space and Astronautical Science (ISAS),
Japan Aerospace Exploration Agency (JAXA), Sagamihara, Kanagawa 252-5210, Japan}

\author{Makoto \textsc{Watanabe}, Masataka \textsc{Imai}, Shuhei \textsc{Goda}}
\affil{Department of Cosmosciences, Graduate School of Science, Hokkaido University,
Sapporo 060-0810, Japan}

\author{Hiroshi \textsc{Akitaya}}
\affil{Hiroshima Astrophysical Science Center, Hiroshima University,
Higashihiroshima, Hiroshima 739-8526, Japan}

\author{Yuhei \textsc{Takagi}, Kumiko \textsc{Morihana}, Satoshi \textsc{Honda}, Akira \textsc{Arai}}
\affil{Nishi-Harima Astronomical Observatory, Center for Astronomy,
University of Hyogo, Sayo, Hyogo 679-5313, Japan}

\author{Kazuhiro \textsc{Sekiguchi}}
\affil{National Astronomical Observatory of Japan, National Institute of Natural
Sciences, Mitaka, Tokyo 181-8588, Japan}

\author{Yumiko \textsc{Oasa}}
\affil{Faculty of Education, Saitama University, Sakura, Saitama
338-8570, Japan}

\author{Yoshihiko \textsc{Saito}}
\affil{Department of Physics, Tokyo Institute of Technology,
Meguro-ku, Tokyo 152-8551, Japan}

\author{Tomoki \textsc{Morokuma}}
\affil{Institute of Astronomy, Graduate School of Science, The University of Tokyo,
Mitaka, Tokyo 181-0015, Japan}

\author{Katsuhiro \textsc{Murata}}
\affil{Department of Astrophysics, Nagoya University, Chikusa-ku, Nagoya 464-8602, Japan}

\author{Daisaku \textsc{Nogami}}
\affil{Department of Astronomy, Graduate School of Science, Kyoto University,
Kyoto 606-8502, Japan}

\author{Takahiro \textsc{Nagayama}}
\affil{Graduate School of Science and Engineering, Kagoshima University,
Kagoshima 890-0065, Japan}

\author{Kenshi \textsc{Yanagisawa}}
\affil{Okayama Astrophysical Observatory, National Astronomical
Observatory of Japan, Asaguchi, Okayama 719-0232, Japan}

\author{Michitoshi \textsc{Yoshida}}
\affil{Hiroshima Astrophysical Science Center, Hiroshima University,
1-3-1 Kagamiyama, Higashi-Hiroshima, Hiroshima 739-8526, Japan} 

\author{Kouji \textsc{Ohta}}
\affil{Department of Astronomy, Kyoto University, Kyoto 606-8502, Japan}

\author{Nobuyuki \textsc{Kawai}}
\affil{Department of Physics, Tokyo Institute of Technology,
Meguro-ku, Tokyo 152-8551, Japan}

\author{Takeshi \textsc{Miyaji}}
\affil{Ishigakijima Astronomical Observatory, National Astronomical
Observatory of Japan, Ishigaki, Okinawa 907-0024, Japan}

\author{Hideo \textsc{Fukushima}, Jun-ichi \textsc{Watanabe}}
\affil{National Astronomical Observatory of Japan, Mitaka, Tokyo
181-8588, Japan} 

\author{Cyrielle \textsc{Opitom}, Emmanu\"{e}l \textsc{Jehin}, Michael \textsc{Gillon}}
\affil{Institut d'Astrophysique de l'Universit$\acute{e}$ de Li$\grave{e}$ge,
All$\acute{e}$e du 6 Ao$\hat{u}$t 17, B-4000 Li$\grave{e}$ge, Belgium}

\author{Jeremie J. \textsc{Vaubaillon}}
\affil{Observatoire de Paris, I.M.C.C.E., Denfert Rochereau, Bat. A.,
FR-75014 Paris, France}

\altaffiltext{\dag}{Visiting Astronomer, Observatoire de Paris, I.M.C.C.E., 
Denfert Rochereau, Bat. A., FR-75014 Paris, France, in 2014 May--July}

\begin{abstract}
We report a new observation of the Jupiter-family comet 209P/LINEAR during its 2014 return.
The comet is recognized as a dust source of a new meteor shower, the May Camelopardalids.
209P/LINEAR was  apparently inactive at a heliocentric distance $r_h$ = 1.6 au and
showed weak activity at $r_h\leq$ 1.4 au.
We found an active region of $<$0.001\% of the entire nuclear surface during the comet's dormant phase.
An edge-on image suggests that particles up to 1 cm in size (with an uncertainty of factor 3--5)
were ejected following a differential power-law size distribution with index $q=-3.25\pm0.10$.
We derived a mass loss rate of 2--10 kg sec$^{-1}$ during the active phase and a total mass of
$\approx$5 $\times$ 10$^7$ kg during the 2014 return.
The ejection terminal velocity of millimeter- to centimeter-sized particles was 1--4 m sec$^{-1}$,
which is comparable to the escape velocity from the nucleus (1.4 m sec$^{-1}$). These results imply
that such large meteoric particles marginally escaped from the highly dormant comet nucleus via the gas
drag force only within a few months of the perihelion passage.
 \end{abstract}

\keywords{comets: individual (209P/LINEAR) --- interplanetary medium --- meteorites, meteors, meteoroids}

\section{Introduction}
\label{sec:introduction}
%Comets are icy leftovers from planetary accretion occurring beyond the snowline
%and lose their mass after they are injected into the inner Solar System. On the other hand,
%meteor showers are caused by streams of dust particles, mostly originating from comets.
The link between comets and meteor showers is important for better
understanding of how pristine cometary materials have been delivered to the Earth.
209P/LINEAR (hereafter 209P) has an orbit typical of Jupiter-family comets,
that is, a semimajor axis $a$=2.932 au, eccentricity $e$=0.692, inclination
$i$=19.4\arcdeg, and Tisserand parameter with respect to Jupiter,
$T_\mathrm{J}$, of 2.80. It was suggested that a swarm of dust from 209P
might cause a meteor shower on UT 2014 May 24 \citep{Jenniskens2014}. 
\citet{Ye2014} has reported that 209P is relatively depleted in dust production, with a low level
of activity around the perihelion passage in 2008--2009.
This paper attempts to characterize the physical properties further through a new observation in 2014.
We focus on the dust ejection properties (e.g., particle size and ejection terminal velocity),
which are pivotal for linking the comet with the meteor shower via a dynamical model \citep[see, e.g.,][]{Vaubaillon2005}.

\section{Observations and Date Analysis}
\label{sec:observation}
The journal of these observations is summarized in Table \ref{tab:journal}.
The first imaging observation was conducted on UT 2014 February 1 using a Tektronix 2048 $\times$ 2048
pixel CCD camera (Tek2k) on the University of Hawaii 2.24-m telescope (UH2.2m) atop Mauna Kea.
We obtained optical images with a broadband Kron--Cousins R$_\mathrm{C}$-band filter. We noticed
that the comet appeared point-like even at a heliocentric distance $r_h$ = 1.57 au, where comets generally
display comae and tails. Later, we made a network observation through the Optical
and Infrared Synergetic Telescopes for Education and Research ({\it OISTER}), which is an inter-university
observation network in the optical and infrared wavelengths.
% that is used to promote education and research at associated universities and institutes in Japan.
Among the {\it OISTER}  network,
we used four telescopes for the present study: the Nishi-Harima Astronomical Observatory
Nayuta 2.0-m telescope (NHAO2m), the Ishigakijima Astronomical Observatory Murikabushi
1.0-m telescope (IAO1m), the Okayama Astrophysical Observatory 0.5-m reflecting telescope
(OAO0.5m), and the Nayoro Observatory 1.6-m Pirka telescope of the Hokkaido University (NO1.6m).
We employed the optical imaging cameras MINT (a back--illuminated 2048 $\times$ 2064 CCD
chip with a 15-\micron\  pixel pitch) with R$_\mathrm{C}$- and  I$_\mathrm{C}$-band filters at NHAO2m,
two sets of MITSuME (g$'$, R$_\mathrm{C}$, and I$_\mathrm{C}$-band simultaneous imaging system,
a 1024 $\times$ 1024 CCD chip with a 24.0-\micron\ pixel pitch) at IAO1m and OAO0.5m, and
the MSI (a visible multispectral imager with a 512 $\times$ 512 CCD chip with a 16.0-\micron\ pixel pitch\citep{Watanabe2012} at NO1.6m.
The two sets of MITSuME at IAO1m and OAO0.5m were designed identically, and each houses three front-illuminated CCD cameras.
%The fields of view (FOVs) and pixel scales of these instruments are
%7.5\arcmin $\times$ 7.5\arcmin\ (0.22\arcsec pixel$^{-1}$) at UH2.2m,
%11\arcmin $\times$ 11\arcmin\ (0.32\arcsec pixel$^{-1}$) at NHAO2m,
%12.3\arcmin $\times$ 12.3\arcmin\ (0.72\arcsec pixel$^{-1}$) at IAO1m,
%26\arcmin $\times$ 26\arcmin\ (1.53\arcsec pixel$^{-1}$) at OAO0.5m, and
%3.3\arcmin $\times$ 3.3\arcmin\ (0.39\arcsec pixel$^{-1}$) at NO1.6m.
After early June 2014, 209P was unobservable from these observatories, which are located in the 
northern hemisphere. Instead, we observed the comet with the 0.6-m Transiting Planets and
Planetesimals Small Telescope (TRAPPIST0.6m) with a 2048 $\times$ 2048 back-illuminated CCD
chip with a 15-\micron\ pixel pitch \citep{Jehin2011}. It covers 22\arcmin $\times$ 22\arcmin\ with a resolution of 1.3\arcsec\
pixel$^{-1}$ using 2$\times$2 binning. All telescopes were operated in a non-sidereal tracking mode so that the comet
was stationary in the observed frames.
% while background objects were essentially elongated during the individual exposures (typically 3--5 min).

The observed data were analyzed in the standard manner for optical and near-infrared imaging data.
We constructed median-stacked frames using 209P frames or dome flat images to correct for the
effect of the pixel-to-pixel
sensitivity variations across the detectors as well as optical vignetting (what is called the flat field image).
The photometric zero levels were determined using Landolt photometric standard stars \citep{Landolt1992}
for UH88 and NO1.6m data and field stars listed in the  USNO-A2.0 catalog \citep{Monet2003} for the others.
The images observed during a single night were combined to confirm the existence of a dust coma and
further investigate the surface brightness profile of the dust tail (see Section \ref{sec:model}).

\section{Results}

\subsection{Appearance}
\label{sec:appearance}
We found no significant morphological differences between the g$'$-, R$_\mathrm{C}$- and I$_\mathrm{C}$-band
images taken with MITSuME. The obtained color indices, $g'-R_\mathrm{C}$=0.8$\pm$0.3 and
$R_\mathrm{C}-I_\mathrm{C}$=0.5$\pm$0.3, are consistent with those of the Sun, that is,
($g'-R_\mathrm{C}$)$_\odot$=0.65 \citep{Kim2012} and ($R_\mathrm{C}-I_\mathrm{C}$)$_\odot$=0.33 \citep{Holmberg2006},
which implies that the reflected light from the nucleus and dust are the dominant light sources of the detected
intensity. In addition, it is reported that the spectrum
taken with the 8-m Gemini North telescope on April 9.25 UT did not reveal obvious emission lines attributable
to sources such as C$_2$ around 4500--5600\AA\ and NH$_2$ around 4900--6300\AA\ \citep{Schleicher2014}.
For these reasons, we ignored the contribution of gaseous emission in our R$_\mathrm{C}$-band data and used
the R$_\mathrm{C}$-band magnitudes for the subsequent photometric analysis (see also Table \ref{tab:journal}).

Figure 1 shows selected R$_\mathrm{C}$-band images of 209P. 
In the first image, taken on UT 2014 February 1 (at $r_h$=1.57 au), neither the coma
nor the dust tail was visually apparent. An unclear tail-like feature extended to the position angle
(the angle on the celestial plane measured from north through east)
PA$\sim$185\arcdeg. It is not clear whether the feature was attributable to the cometary tail or
an artifact such as a diffraction spike from the support vanes of the secondary mirror.
The 1.05\arcsec--1.06\arcsec\ full width at half-maximum (FWHM) of the field stars is in perfect agreement
with the value of 1.05\arcsec\ in the combined 209P image. In Figure 2 (a), we compare the radial profile of 209P
in a composite image with that of a field star taken in sidereal tracking mode between the 209P exposures.
We found that the surface brightness profiles coincided with one another at the 10$^{-3}$--
10$^{-2}$\% level of the photocenter. The similarity suggests that the comet was highly
dormant on that night (UT 2014 February 01). We set an upper limit of 0.01 for the parameter $\eta$, which is
defined as the ratio of the coma cross section to the nucleus cross section. Adopting a model in \citet{luu1992}
and assuming the ejection of small dust particles (a radius of $a_\mathrm{d}$=0.5 \micron) that are embedded
in surface water ice, we obtained approximate estimates
for the dust production rate $M_\mathrm{d}\lesssim$0.01 kg s$^{-1}$ and the fractional active area
$f\lesssim$1 $\times$ 10$^{-5}$ on UT 2014 February 1 \citep[see also][]{Ishiguro2011},
The obtained $f$ value is significantly lower than those of the typical Jupiter-family comets \citep[$f>$10$^{-3}$,][]{Tancredi2006}.

In  Figure 1 (b) (UT 2013 March 03 at $r_h$=1.30 au), the comet still appeared point-like.
However, a careful investigation revealed a faint tail-like structure extending to 
PA=128$\pm$3\arcdeg, which is close to the position angles of the Sun--comet radius vector (PA=123\arcdeg)
but deviates slightly to the negative heliocentric velocity vector (PA=216\arcdeg). Since cometary dust tails
usually appear between these two vectors, and the position angle does not align with the diffraction spike
caused by the secondary mirror, we suspect that the extended structure might be a real cometary tail.
In  Figure 1 (c) (UT 2013 March 23), the cometary tail was clearly detected. 
It extended to PA=105$\pm$4\arcdeg, existing between the antisolar direction (PA=100\arcdeg) and
the negative heliocentric velocity vector (PA=193\arcdeg).
We detected an obvious tail  in all the images after UT 2013 March 22.
Figure 1 (d) was taken when the comet was viewed edge-on on UT 2013 May 23.
Note that the image was rotated to align the projected orbital plane in the horizontal direction.
The comet possessed a narrow tail extended to PA=108$\pm$1\arcdeg, which coincided with the position angle
of the orbital plane projected on the sky (PA=107.4\arcdeg). The tail extended out of the FOV (i.e., $>$13\arcsec).
Further, the dust cloud extended sunward by 30\arcsec\ (rightward in the image), probably because of the ejection of fresh
dust particles toward the Sun. To obtain a crude estimate of the ejection velocity, we employed the formula
$l=v_{ej}^2/(2\beta g_\odot)$, where $l$ is the apparent length of the sunward tail, $v_{ej}$ is the terminal escape velocity
of dust particles, $\beta$ is the ratio of the solar radiation pressure to the solar gravity, and $g_\odot$ is the solar gravity
at the position of the comet \citep{Jewitt1987}. We obtained 1.1 m sec$^{-1}$ assuming 1-cm particles and
3.4 m sec$^{-1}$ assuming 1-mm particles. The order of magnitude estimate for $v_{ej}$ is consistent with the result
of another model described below (Section \ref{sec:model}).

\subsection{Properties of Nucleus}
\label{sec:nucleus}
Figure 2 (b) shows the lightcurves of 209P measured from each image on UT 2014 February 1.
The data were calibrated using Landolt photometric standard stars, ensuring an absolute magnitude
accuracy of 0.05 mag or less \citep{Landolt1992}. The rotational lightcurve covered one peak and probably two troughs
(both ends), suggesting that the rotational period is not shorter than the observational duration (7 h).
The inferred rotational period is consistent with a report by Hergenrother in which he derived two alternative solutions of
10.930$\pm$0.015 and 21.86$\pm$0.04 h \citep{Green2014}. 
%The lightcurve shows an amplitude of $\sim$0.35 at the phase angle (sun--comet--observer's angle) $\alpha$=27.6\arcdeg.
We calculated the corresponding amplitude at $\alpha$ = 0\arcdeg\ using an empirical function \citep{Zappala1990},
\begin{eqnarray}
A\left(0\right)=\frac{A\left(\alpha\right)}{1+m\alpha},
\label{eq:zappala}
\end{eqnarray}
\noindent where $A\left(0^\circ\right)$ and $A\left(\alpha\right)$ are the amplitudes at phase angles (Sun--comet--observer angles)
of 0\arcdeg\ and $\alpha$, respectively, and $m$ is a correction coefficient for the amplitude, which has different values
for S-, C-, and M-type asteroids. We adopted $m$ = 0.015, the value for C-type asteroids, because the comet nucleus may have
optical properties similar to those of C-type asteroids rather than S- or M-type asteroids. Substituting $m$ = 0.015 and
$\alpha=$ 27.6\arcdeg, we obtained an axis ratio of 1:1.25. 
%Together with the shape and rotational period, we derived
%a reasonable lower limit for the mass density, 0.06 g cm$^{-3}$, by applying a formula in \citet{Harris1996}.

The magnitude is related to the effective (or mean) radius of the nucleus, $r_n$, by
\begin{eqnarray}
p_\mathrm{R}\Phi\left(\alpha\right)r_n^2=2.25\times10^{22}r_h^2\Delta^210^{-0.4\left(m_\mathrm{R}-m_\odot\right),}
\label{eq:zappala}
\end{eqnarray}

\noindent where $p_\mathrm{R}$ is the geometric albedo in the R$_\mathrm{C}$ band; $\Phi\left(\alpha\right)$
is the phase function; $r_h$ and $\Delta$ are the heliocentric and geocentric distances, respectively, in au;
and $m_\odot=-27.1$ is the apparent R$_\mathrm{C}$ magnitude of the Sun. $\Phi\left(\alpha\right)$ is
often assumed to be $\Phi\left(\alpha\right)$ = 10$^{-0.4 b\alpha}$, where $b$ is a parameter characterizing
the phase slope \citep{Belskaya2000}. We assumed $b$ = 0.04 mag deg$^{-1}$ and  $p_\mathrm{R}$ = 0.05,
and obtained the R$_\mathrm{C}$ band absolute magnitude H$_\mathrm{R}$=16.24 and $r_n$=1.4 km,
or the dimension of 2.5 $\times$ 3.2 km. Although there are uncertainties in
$b$ (from 0.035 to 0.045 mag deg$^{-1}$, Belskaya et al. 2000) and $p_\mathrm{R}$ (from 0.03 to 0.07, Kim et al. 2014),
which cause a 40\% error ($\sim$1 km) in the size, the derived size is in good agreement with that determined by a radar observation
\footnote{http://www.usra.edu/news/pr/2014/comet209PLINEAR/}, which reported dimensions of 2.4 $\times$ 3.0 km.
The similarity may suggest that the comet was inactive on 2014 February 1 and has optical properties typical of
comet nuclei. For comparison, We fit our data at low phase angle ($\alpha<$40.8\arcdeg) using the $H$--$G$ formalism
\citep{lumme1984,Bowell1989} when the comet was apparently inactive,
and obtained H$_\mathrm{R}$=16.11$\pm$0.26 and G=0.15$\pm$0.17. 

\subsection{Coma Photometry}
\label{sec:coma}
Figure 3 (a) shows the  R$_\mathrm{C}$-band reduced magnitude (a hypothetical magnitude
observed at 1 au from both the Earth and the Sun) with respect to the phase angle.
We set the aperture size for photometry to 3 times the FWHM of point sources (5\arcsec--9\arcsec, depending on
the sky conditions). In the figure, we considered the uncertainty of the magnitude on the basis of two factors: one is associated with
the uncertainty of the magnitudes of comparison stars (0.25 mag for USNO-A2.0), and the other results from the rotation
of the nucleus (a half amplitude of the lightcurve, 0.18 mag), because most of our data could not cover an adequate
rotational phase (except the data from UH2.2m and NO1.6m).
%The photon and readout noise are negligible compared to these two factors.
The data taken on UT 2014 February 01 ($\alpha=$ 27.6\arcdeg) have the smallest error not only because they were calibrated
with appropriate standard stars in the Landolt catalog, but also because the data covered a substantial rotational phase for deriving the mean magnitude.
In Figure 3 (a), we show the reference magnitude of the nucleus, which is given by
$m_R\left(\alpha\right)=16.24+0.04\alpha$, following the result in Section \ref{sec:nucleus}.
The magnitude of 209P was significantly brighter than
the predicted nuclear magnitude at $\alpha\gtrsim$ 50\arcdeg.
Since we considered the rotational brightening/darkening in the error bars, it is unlikely that the magnitude enhancement was
caused by sampling bias.
When we force fitted the magnitude data with a linear function, we obtained a
phase slope of $b$ = 0.03, which is inconsistent with low-albedo objects \citep[see][]{Belskaya2000}. Therefore, it is reasonable
to think that the magnitude enhancement was caused by a dusty coma near the nucleus.

Figure 3 (b) and (c) show the differences in magnitude between the observation and the nucleus model
with respect to true anomaly $\theta_T$ and the heliocentric distance $r_h$, respectively.
There seems to be a weak trend that the residual increased toward perihelion [see Figure 3 (c)].
The magnitude enhancement appears at $r_h=$ 1.2--1.4 au (or $\theta_T=$ 285--300\arcdeg),
although the tail was not obvious in our composite images.
We conjecture that the nuclear magnitude was brightened at $r_h=$ 1.2--1.4 au
because of a thin dusty coma, although it was not noticeable in our images.
It is thus likely that the tail-like feature in Figure 1 (b) could be a dust tail associated with weak comet-like activity
(see \ref{sec:appearance}). We also noticed that the comet's activity may not be symmetric with respect to perihelion.
The differential magnitude has a peak at $\theta_T \sim$340\arcdeg, which is close to perihelion but
slightly shifted toward the inbound orbit.
Generally, activity peaks of comets tend to shift toward the post-perihelion passages \citep[see, e.g.,][]{Ferrin2010}.
We conjecture that the activity peak prior to the perihelion may be associated with the seasonal variation of solar incident flux at a localized
active region, as indicated for 9P/Tempel 1 \citep{Schleicher2007}.

\subsection{Dust Tail and Meteoroid Ejection}
\label{sec:model}
To link a comet with a meteor shower, it is important to know how meteoric particles were ejected from the nucleus.
We determine the size and ejection velocity using a simple but straightforward method shown below.

We noticed that the edge-on image provides a unique opportunity for deriving the size and ejection velocity.
It was taken on UT 2014 May 23 [Figure 1 (d)] in a nearly edge-on view;
that is, the angle between the observer and the 209P orbital plane was 3\arcdeg. 
Figure 4 shows the surface brightness
($\Sigma$) profile of the dust tail integrated within a width of 3\arcmin\ perpendicular to the projected orbit,
as a function of the distance from the nucleus, $d$.
The profile at $d\lesssim$ 6\arcsec\ was contaminated by light from the nucleus.
Since the comet moved rapidly on the sky plan (12\arcmin\ min$^{-1}$), it was
elongated up to 8\arcsec\ by inadequate tracking of the telescope.

In Figure 4, we found that an inflection point exists at $d\sim50$\arcsec. 
The surface brightness along the tail is consistent with $\Sigma \propto d^\gamma$,
where $\gamma=-0.57\pm0.05$ at $d$ = 10\arcsec--50\arcsec\ and $\gamma=-0.73\pm0.03$ at
$d$ = 50\arcsec--300\arcsec. Because the difference in $\gamma$ is significant to the accuracy of 
our measurement, we attribute the discontinuity at $d\sim50$\arcsec\ to a discontinuous
distribution of dust particles. When dust particles are ejected at a constant rate over a long interval,
the resulting steady-state flow of dust particles yields a surface brightness distribution with $\gamma$ = $-$0.5.
The similarity in $\gamma$ values between the observed data at $d$ = 10\arcsec--50\arcsec\ and a steady-state
flow suggests that dust particles flowed steadily owing to solar radiation pressure near the nucleus ($d<50$\arcsec). 
In contrast, the steeper slope beyond $d=50$\arcsec\ may suggest that only smaller particles reach the region,
as considered in \citet{Jewitt2014}. Assuming that the dust particles were ejected after late March at a constant rate,
$d<50$\arcsec\ corresponds to $\beta>$ 3 $\times$ 10$^{-5}$ or $a\lesssim$ 1 cm (a density of
$\rho$=1 g cm$^{-3}$ is assumed), where $\beta$ is again the ratio of  the solar radiation pressure
acceleration to solar gravity. We adopted the continuous dust ejection model in \citet{Jewitt2014} 
and found that the dust particles have a differential power-law size distribution with index $q\sim$ 3.25$\pm$0.10.
The ejection velocity perpendicular to the orbital plane was 0.7 m sec$^{-1}$ for 1-cm grains.
Assuming that dust particles were ejected symmetrically to the comet--Sun vector within a half opening angle
of 30--60\arcdeg, the net ejection velocity is estimated to be 0.8--1.4 m s$^{-1}$.
With the model, we also estimated the ejection velocity of 1-mm particles as 2.5--4.4 m s$^{-1}$.
The velocity is consistent with or slightly faster than the escape velocity (1.4 m s$^{-1}$) from an 1850-m body
with a nuclear mass density of 1 g cm$^{-3}$. Assuming that the dust has the same optical properties as the nucleus,
we derived a total dust grain mass of (2--8) $\times$ 10$^7$ kg.
Assuming that the particles were ejected for three months, from late March until late May,
we obtained an average mass loss rate around perihelion of 2--10 kg s$^{-1}$.
The model predicts a loss of 2 $\times$ 10$^8$ particles s$^{-1}$ for $>$1-mm particles.
There seems to be an uncertainty of 3--5 times in the particle size due to the uncertain onset time of the active phase (i.e., late February
or late March) and mass density (0.3--2 g cm$^{-3}$). The uncertainty is translated into an uncertainty
of 3--5 in the particle production rate.
Considering all of the results above, we concluded that meteoric particles (1--10 mm) were marginally ejected
from the highly dormant comet nucleus via gas outflow only when the comet was around perihelion.

The peak activity of the Camelopardalids occurred on UT 2014 May 24 as predicted. \citet{Brown2014} reported that
the shower signals were dominated by small particles of milligram mass and smaller (i.e. $\lesssim$1 mm).
Further research is needed to connect the observed mass ejection for 209P and meteor shower, taking account
of dynamical evolution \citep{Vaubaillon2005} and probably fragmentation of dust aggregates \citep{Madiedo2014}.

\section{Summary}
\label{sec:summary}
We made observations of 209P during its perihelion passage in 2014 and found the following:
\begin{enumerate}

\item{209P/LINEAR was  apparently inactive at the heliocentric distance $r_h$ = 1.6 au and
showed weak activity at $r_h\leq$ 1.4 au.}
\item{The observed morphology is similar in the R$_\mathrm{C}$ and I$_\mathrm{C}$ bands, suggesting
that scattered sunlight from the nucleus and dust particles was the dominant light source at these optical wavelengths.}
\item{The comet was determined to have a negligibly small active fraction ($<$0.001\%) based on upper limit coma measurements
made prior to the appearance of clear cometary activity.}
\item{During the active phase, it ejected dust particles up to 1 cm in size with a differential power-law size distribution with index $q=-3.25\pm0.10$.}
\item{The total ejected dust mass and average mass loss rate were (2--8) $\times$ 10$^7$ kg and 2--10 kg sec$^{-1}$, respectively.}
\end{enumerate}

\vspace{1cm}
{\bf Acknowledgments}\\
This research was conducted as part of a joint research project titled ``Recherche sur les liens entre com\`etes et m\'et\'eores.''
MI was supported by the Paris Observatory during his stay in Paris (2014 May--July). This research was also supported
by a National Research Foundation of Korea (NRF) grant funded by the Korean government (MEST) (No. 2012R1A4A1028713). 
The observations at OAO, IAO, NHAO, and NO were supported by the Optical and Near-infrared Astronomy Inter-University
Cooperation Program and Grants-in-Aid for Scientific Research (23340048, 24000004, 24244014, and 24840031)
from the Ministry of Education, Culture, Sports, Science and Technology of Japan.
TRAPPIST is a project funded by the Belgian Fund for Scientific Research (Fonds National de la Recherche Scientifique, F.R.S.-FNRS).
C. Opitom acknowledges the support of the FNRS. E. Jehin and M. Gillon are FNRS Research Associates.
SH was supported by the Space Plasma Laboratory, ISAS, JAXA.

%%%%%%%%%%%
%%   FIGURE 21  %%
%%%%%%%%%%%
\clearpage
\begin{figure}
\epsscale{0.75}
\plotone{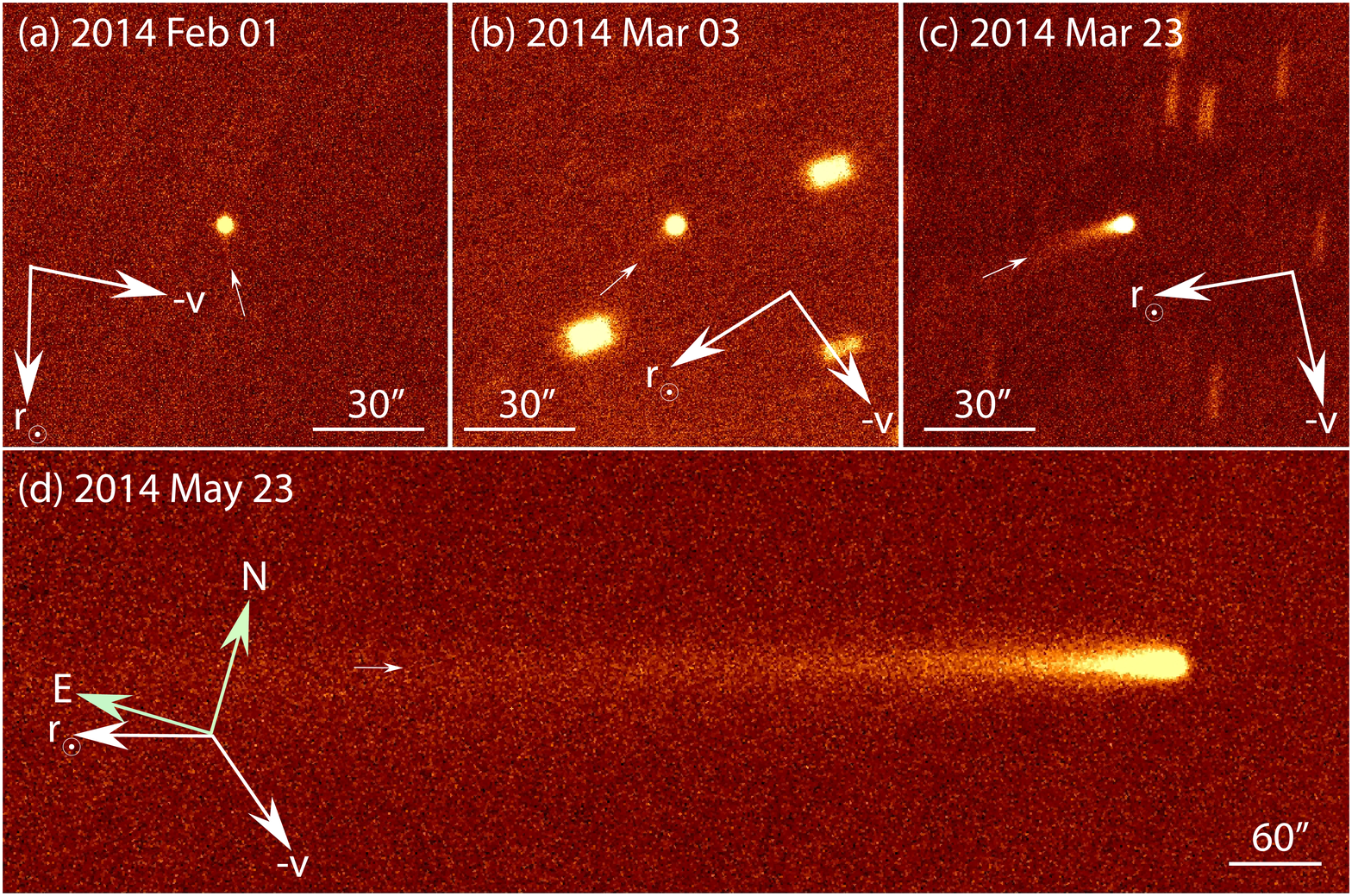}
\caption{Selected images of 209P.
% taken on UT 2014 February 1, March 03, March 23, and May 23.
 The top three images (a--c) have the standard orientation in the sky:
north is up, and east is to the left, and the bottom image (d) is rotated by -17\arcdeg\ so that the Sun--comet
vector is parallel to the horizontal axis. The FOV is 2\arcmin $\times$ 2\arcmin (a--c) and
14.5\arcmin $\times$ 4.8\arcmin (d). 
%Background stars and galaxies were removed
%from the original data, although several bright stars were not erased properly in images on March 03 and March 23.
The antisolar direction ($r_\odot$) and the negative heliocentric velocity vector ($-v$) are shown by arrows.
Thin arrows indicate possible dust tails.
\label{fig2}}
\end{figure}

%%%%%%%%%%%
%%   FIGURE 1   %%
%%%%%%%%%%%
\clearpage
\begin{figure}
\epsscale{0.75}
\plotone{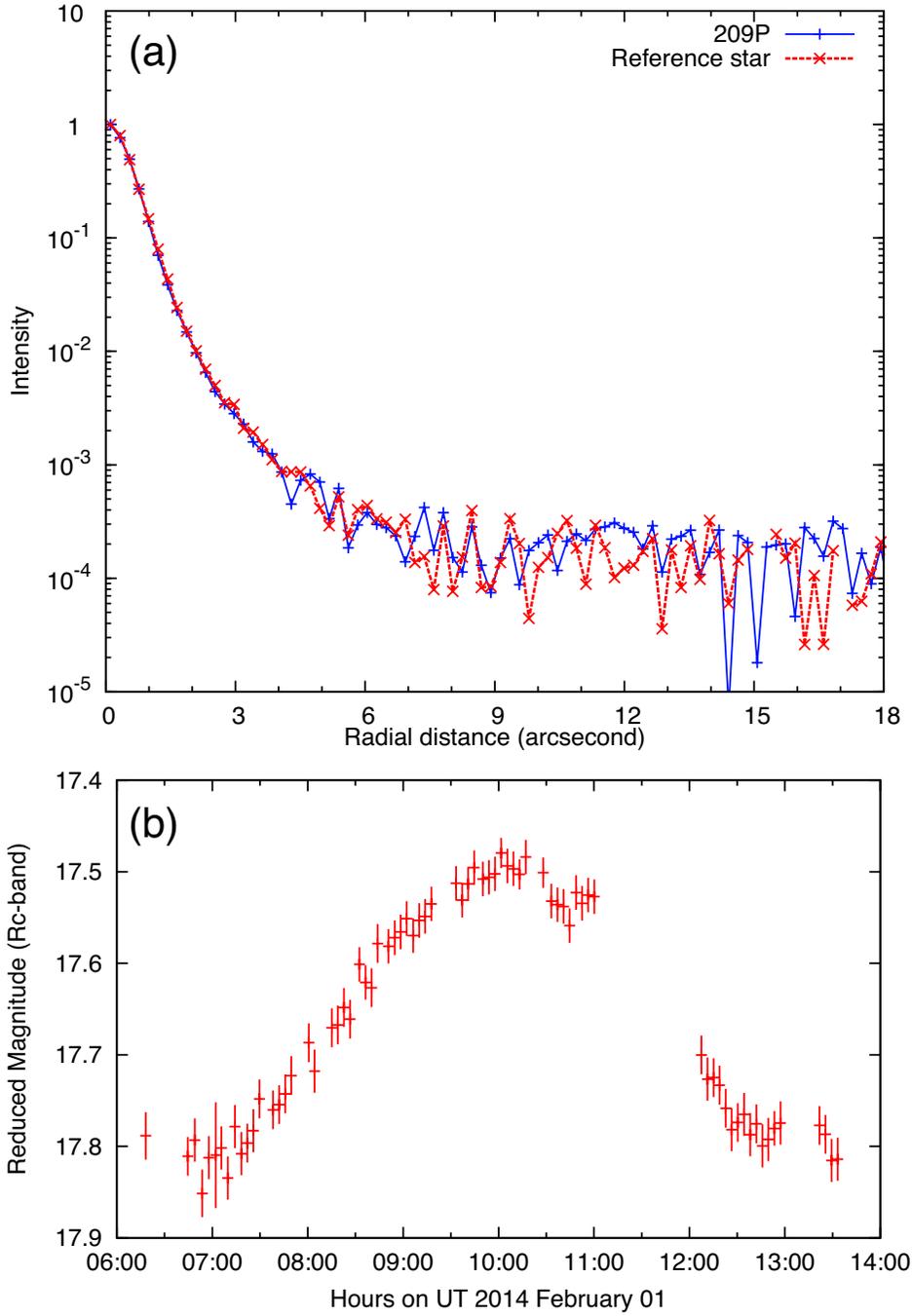}
\caption{%Surface brightness and rotational magnitude of 209P.
(a) Normalized surface brightness profiles of 209P (solid line) and a reference star (dashed line) taken on
UT 2014 February 1. The stellar profile was taken in sidereal tracking mode six times at the beginning, middle and end
of 209P exposures with the exposure time of 180 seconds. We could not find a noticeable time-variation in the stellar profiles.
(b) Rotational lightcurve on the same night. Vertical axis denotes the reduced magnitude,
and horizontal axis denotes UT on 2014 February 01 after light time correction.
\label{fig:f1}}
\end{figure}

%%%%%%%%%%%
%%   FIGURE 3   %%
%%%%%%%%%%%
\clearpage
\begin{figure}
\epsscale{0.75}
\plotone{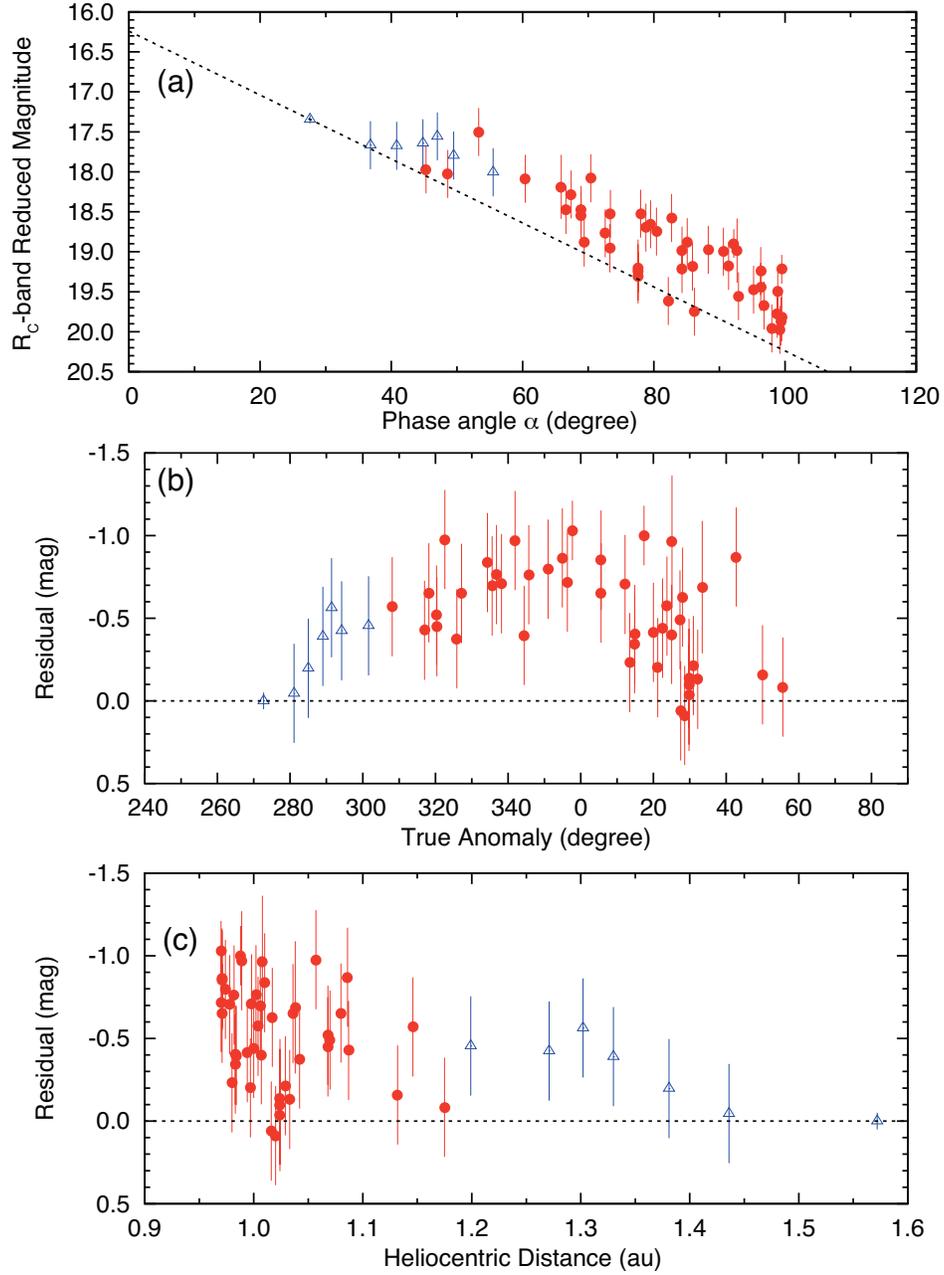}
\caption{ Photometric results: (a) Magnitude--phase relation of 209P/LINEAR. Dashed line denotes the
predicted mean magnitude of the rotating nucleus. (b) Residual of magnitudes after subtraction
of the nuclear contribution with respect to the true anomaly $\theta_{T}$. (c) Residual of magnitudes
with respect to the distance from the Sun. Open triangles are magnitudes when the comet showed
obvious tail while filled circles are magnitude when the comet appeared point-like.
%We set the aperture sizes for photometry to 3 times the FWHM, (5--9\arcsec). 
\label{fig:f3}}
\end{figure}

%%%%%%%%%%%
%%   FIGURE 4   %%
%%%%%%%%%%%
\clearpage
\begin{figure}
\epsscale{1.0}
\plotone{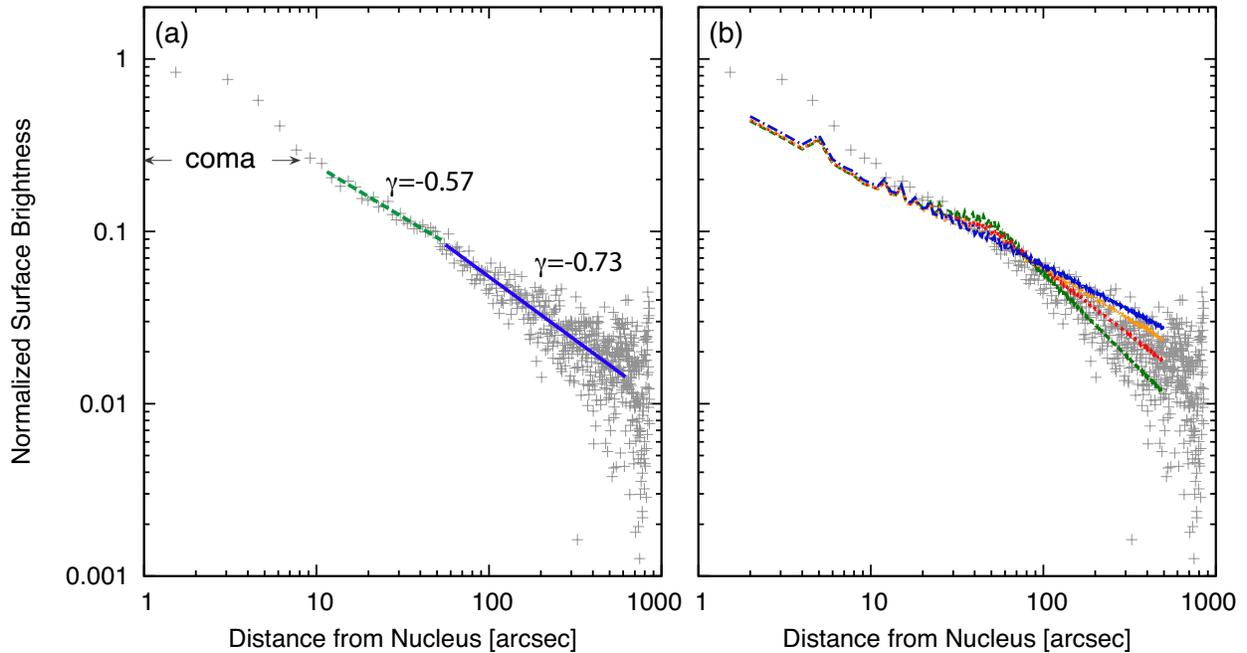}
\caption{Surface brightness profiles of 209P (crosses) with respect to distance from the nucleus
observed on UT 2014 May 23. (a) The profile was fitted by power-law functions with indexes
$\gamma$ = $-$0.57 ($d$ = 10\arcsec--50\arcsec) and $\gamma=-0.73\pm0.03$ ($d$ = 50\arcsec--300\arcsec).
(b) Model profiles in which dust is ejected continuously starting on UT 2014 February 22.
We assumed the minimum $\beta$ (= 3 $\times$ 10$^{-5}$), which corresponds
to 1-cm grains, to produce the observed inflection point at $d\sim$ 50\arcsec.
The power-law indices are $q=-3.75$ (blue), $-3.50$ (orange), $-3.25$ (red), and $-3.00$ (green) from top to bottom.
\label{fig:f4}}
\end{figure}

%\clearpage
%\begin{table}
%\begin{center}
%\footnotesize
%\caption{Summary of Telescopes and Instruments\label{tab:instruments}}
%\begin{tabular}{ccccccc}
%\tableline\tableline
%Observatory & Telescope & Instruments & Modes & Field-of-view & Pixel scale\\
%\tableline
%Hokkaido University & 1.6-m Pirika & MSI & Photopolarimetry & 3.3\arcmin$\times$3.3\arcmin\\
%UH & 2.2-m & Tek2k & Optical photometry & 7.5\arcmin$\times$7.5\arcmin & 0.22\arcsec\\
% NHAO & 2.0-m Nayuta & MINT & Optical photometry & 11\arcmin$\times$11\arcmin & 0.32\arcsec\\
% & & NIC & Near-IR photometry & 2.7\arcmin$\times$2.7\arcmin & 0.16\arcsec\\
%NAOJ/OAO & 0.5-m Reflector & MITSuME & Optical photometry & 26\arcmin$\times$26\arcmin & 1.53\arcsec\\
%Hiroshima University & 1.5-m Kanata & HOWPol & Photopolarimetry & 15\arcmin circular\\
%NAOJ/IAO & 1.0-m Murikabushi & MITSuME & Optical photometry & 12\arcmin$\times$12\arcmin & 0.72\arcsec\\
%\tableline
%\end{tabular}
%\end{center}
%\end{table}
%
%

\clearpage
\begin{deluxetable}{lcccccccccc}
\tablecaption{Observation summary\label{tab:journal}}
\tablewidth{375pt}
\tabletypesize{\scriptsize}
%\tablenum{2}
\tablehead{\colhead{Median UT} & \colhead{Telescope} & \colhead{Filter} & \colhead{$N^a$} & \colhead{$T_{tot}^b$} & \colhead{r$_h^c$} & \colhead{$\Delta^d$} & \colhead{$\alpha^e$} & \colhead{$f_T^f$} & \colhead{Mag$^g$} & \colhead{Tail$^h$}}
\startdata
2014-Feb-01.418 & UH 2.2 m &R$_\mathrm{C}$&72&216&1.572&0.729&27.6&272.7&17.6 & No\\
2014-Feb-16.540 & IAO 1.0 m &g$'$, R$_\mathrm{C}$, I$_\mathrm{C}$&17&51&1.436&0.663&36.8&281.1&17.6 & No\\
2014-Feb-22.697& IAO 1.0 m &g$'$, R$_\mathrm{C}$, I$_\mathrm{C}$&17&51&1.381&0.641&40.8&285.0&17.4 & No\\
2014-Feb-28.598 & IAO 1.0 m &g$'$, R$_\mathrm{C}$, I$_\mathrm{C}$&19&57&1.330&0.622&44.8&289.0&17.2 & No\\
2014-Mar-03.855 & NHAO 2.0 m &R$_\mathrm{C}$&20&10&1.302&0.611&47.0&291.4&17.1 & No\\
2014-Mar-07.641 & NHAO 2.0 m &R$_\mathrm{C}$&55&27.5&1.271&0.598&49.5&294.2&17.2 &No\\
2014-Mar-16.551 & IAO 1.0 m &g$'$, R$_\mathrm{C}$, I$_\mathrm{C}$&11&33&1.199&0.565&55.5&301.6&17.2 &No\\
2014-Mar-22.604 & IAO 1.0 m &g$'$, R$_\mathrm{C}$, I$_\mathrm{C}$&24&72&1.154&0.539&59.6&307.1 & -- &Yes\\
2014-Mar-23.660 & NHAO 2.0 m &R$_\mathrm{C}$&45&90&1.146&0.534&60.4&308.1&17.0& Yes\\
2014-Apr-01.502 & OAO 0.5 m &g$'$, R$_\mathrm{C}$, I$_\mathrm{C}$&53&53&1.087&0.488&66.6&317.0&17.1& Yes\\
2014-Apr-02.618 & NHAO 2.0 m &R$_\mathrm{C}$&14&28&1.080&0.481&67.4&318.2&16.9& Yes\\
2014-Apr-04.562 & IAO 1.0 m &g$'$, R$_\mathrm{C}$, I$_\mathrm{C}$&18&54&1.068&0.470&68.9&320.3&17.0& Yes\\
2014-Apr-04.620 & NHAO 2.0 m &R$_\mathrm{C}$&20&40&1.068&0.470&68.9&320.4&17.1& Yes\\
2014-Apr-06.643 & NHAO 2.0 m &R$_\mathrm{C}$&41&82&1.057&0.457&70.4&322.6&16.5& Yes\\
2014-Apr-09.514 & OAO 0.5 m &g$'$, R$_\mathrm{C}$, I$_\mathrm{C}$&55&55&1.042&0.439&72.6&325.9&17.1& Yes\\
2014-Apr-10.596& IAO 1.0 m &g$'$, R$_\mathrm{C}$, I$_\mathrm{C}$&20&60&1.036&0.432&73.4&327.2&16.8& Yes\\
2014-Apr-16.534 & NHAO 2.0 m &R$_\mathrm{C}$&20&40&1.010&0.390&78.0&334.3&16.5& Yes\\
2014-Apr-17.610& IAO 1.0 m &g$'$, R$_\mathrm{C}$, I$_\mathrm{C}$&18&54&1.006&0.382&78.8&335.6&16.6& Yes\\
2014-Apr-18.547 & NHAO 2.0 m &R$_\mathrm{C}$&30&60&1.002&0.375&79.5&336.8&16.5& Yes\\
2014-Apr-19.638 & NHAO 2.0 m &R$_\mathrm{C}$, I$_\mathrm{C}$&30&60&0.998&0.366&80.4&338.2&16.6& Yes\\
2014-Apr-22.549 & NHAO 2.0 m &R$_\mathrm{C}$&32&63&0.989&0.343	&82.7&341.9&16.2& Yes\\
2014-Apr-24.503 & OAO 0.5 m &g$'$, R$_\mathrm{C}$, I$_\mathrm{C}$&36&72&0.984&0.328&84.3&344.4&16.8& Yes\\
2014-Apr-25.544 & NHAO 2.0 m &R$_\mathrm{C}$&40&40&0.982&0.319&85.1&345.7&16.4& Yes\\
2014-Apr-29.572& IAO 1.0 m &g$'$, R$_\mathrm{C}$, I$_\mathrm{C}$&11&33&0.974&0.286&88.3&351.0&16.2& Yes\\
2014-May-02.486 & OAO 0.5 m &g$'$, R$_\mathrm{C}$, I$_\mathrm{C}$&36&72&0.971&0.260&90.6&354.9&16.0& Yes\\
2014-May-03.584 & NHAO 2.0 m &R$_\mathrm{C}$&40&80&0.970&0.251&91.4&356.3&16.1& Yes\\
2014-May-04.610& NO 1.6 m &R$_\mathrm{C}$&11&17&0.970&0.242&92.2&357.7&15.8& Yes\\
2014-May-10.491 & OAO 0.5 m &g$'$, R$_\mathrm{C}$, I$_\mathrm{C}$&36&72&0.971&0.190&96.3&5.6&15.8& Yes\\
2014-May-10.534 & NHAO 2.0 m &R$_\mathrm{C}$&27&40.5&0.971&0.190&96.3&5.6&15.6& Yes\\
2014-May-15.531& NHAO 2.0 m &R$_\mathrm{C}$&11&16.5&0.978&0.145&98.9&12.2&15.3& Yes\\
2014-May-16.542& NHAO 2.0 m &R$_\mathrm{C}$&40&60&0.980&0.136&99.2&13.5&15.6& Yes\\
2014-May-17.486& OAO 0.5 m &g$'$, R$_\mathrm{C}$, I$_\mathrm{C}$&45&45&0.983&0.128&99.4&14.8&15.4& Yes\\
2014-May-17.541& NHAO 2.0 m &R$_\mathrm{C}$&39&58.5&0.983&0.127&99.5&14.9&15.3& Yes\\
2014-May-19.888& NO 1.6 m &R$_\mathrm{C}$&12&8&0.988&0.110&99.5&17.4&14.4& Yes\\
2014-May-21.547& OAO 0.5 m &g$'$, R$_\mathrm{C}$, I$_\mathrm{C}$&193&96.5&0.994&0.094&98.8&20.0&14.6& Yes\\
2014-May-22.508& OAO 0.5 m &g$'$, R$_\mathrm{C}$, I$_\mathrm{C}$&73&73&0.997&0.087&98.0&21.2&14.6& Yes\\
2014-May-23.541& OAO 0.5 m &g$'$, R$_\mathrm{C}$, I$_\mathrm{C}$&167&167&1.000&0.079&96.8&22.5&14.2& Yes\\
2014-May-24.491& OAO 0.5 m &g$'$, R$_\mathrm{C}$, I$_\mathrm{C}$&26&26&1.004&0.073&95.2&23.7&13.8& Yes\\
2014-May-25.504& OAO 0.5 m &g$'$, R$_\mathrm{C}$, I$_\mathrm{C}$&51&51&1.007&0.067&92.9&25.0&13.7& Yes\\
2014-May-25.586& IAO 1.0 m &g$'$, R$_\mathrm{C}$, I$_\mathrm{C}$&11&33&1.008&0.066&92.7&25.1 &13.1& Yes\\
2014-May-27.522& OAO 0.5 m &g$'$, R$_\mathrm{C}$, I$_\mathrm{C}$&230&115&1.070&0.058&85.9&27.3&13.2& Yes\\
2014-May-27.535& NHAO 2.0 m &R$_\mathrm{C}$&92&23.25&1.016&0.058&86.2&27.5 &13.1& Yes\\
2014-May-28.006&TRAPPIST 0.6 m&R$_\mathrm{C}$&2&6&1.017&0.057&84.3&28.0&12.7& Yes\\
2014-May-28.477& OAO 0.5 m &g$'$, R$_\mathrm{C}$, I$_\mathrm{C}$&18&9&1.020&0.056&82.2&28.6 &13.4& Yes\\
2014-May-29.497& OAO 0.5 m &g$'$, R$_\mathrm{C}$, I$_\mathrm{C}$&52&26&1.024&0.055&77.6&29.8 &13.0& Yes\\
2014-May-29.504& IAO 1.0 m &g$'$, R$_\mathrm{C}$, I$_\mathrm{C}$&46&23&1.024&0.056&77.6&29.8 &13.0& Yes\\
2014-May-29.513& NHAO 2.0 m &R$_\mathrm{C}$&105&26.25&1.024&0.056&77.6&29.8 & 13.1& Yes\\
2014-May-30.489& OAO 0.5 m &g$'$, R$_\mathrm{C}$, I$_\mathrm{C}$&102&51&1.029&0.057&73.3&31.0&12.7& Yes\\
2014-May-31.478& OAO 0.5 m &g$'$, R$_\mathrm{C}$, I$_\mathrm{C}$&52&26&1.033&0.059&69.4&32.2&12.8& Yes\\
2014-Jun-01.506& IAO 1.0 m &g$'$, R$_\mathrm{C}$, I$_\mathrm{C}$&7&3.5&1.038&0.064&65.9&33.4 & 12.1& Yes\\
%2014-Jun-06.007&TRAPPIST0.6m&R$_\mathrm{C}$&1.062&0.093&57.1&38.5&12.5& Yes\\
2014-Jun-10.028&TRAPPIST 0.6 m&R$_\mathrm{C}$&1&1&1.086&0.127&53.2&42.9&13.2& Yes\\
2014-Jun-16.990&TRAPPIST 0.6 m&R$_\mathrm{C}$&5&5&1.132&0.191&48.6&50.0&14.7& Yes\\
2014-Jun-24.028&TRAPPIST 0.6 m&R$_\mathrm{C}$&6&6&1.183&0.258&44.8&56.6&15.3& Yes\\
\enddata
\tablenotetext{a}{Number of exposures}
\tablenotetext{b}{Total exposure time (min)}
\tablenotetext{c}{Heliocentric distance (au)}
\tablenotetext{d}{Geocentric distance (au)}
\tablenotetext{e}{Solar phase angle (degree)}
\tablenotetext{f}{True anomaly (degree)}
\tablenotetext{g}{R$_\mathrm{C}$-band magnitudes}
\tablenotetext{h}{Is a tail clearly observed?}

\end{deluxetable}

\end{document}